\documentclass[12pt]{article}
\usepackage{putex}
\usepackage{graphicx}

\begin{document}

\preprint{PUPT-2288}

\institution{PU}{Joseph Henry Laboratories, Princeton University, Princeton, NJ 08544}

\title{Time warps}

\authors{Steven S. Gubser}

\abstract{I reconsider asymmetrically warped compactifications, in which time and space have different warp factors.  I call such compactifications time warps if the bulk geometry has neither entropy nor temperature.  I provide an example starting from an asymptotically $AdS_5$ spacetime where the speed of light, measured in a fixed coordinate system, is larger near the boundary than it is deep in the interior.  This example follows the general plan of earlier work on superconducting black holes.  To obtain a normalizable, four-dimensional graviton, one can introduce a Planck brane whose action includes a wrong-sign Einstein-Hilbert term.  The equation of state of the Planck brane has $w < -1$, which is a violation of the null energy condition.  I show, in an almost dimension-independent fashion, that such a violation must occur in a static time warp geometry.  Time warps of the type I describe provide an extra-dimensional description of boost invariance as an emergent symmetry in the infrared.  High-energy violations of Lorentz symmetry, if confined to a strongly coupled unparticle sector dual to the time warp geometry, might manifest themselves through unusual kinematic constraints.  As an example, I explain how modifications of unparticle phase space would affect the decay of a heavy particle into a light visible sector particle plus unparticle stuff.}

\date{December 2008}

\maketitle
\tableofcontents

\section{Introduction}
\label{INTRODUCTION}

The usual approach to extra dimensions in string theory is to consider a direct product,
 \eqn{Direct}{
  ds_{10}^2 = -dt^2 + d\vec{x}^2 + d\tilde{s}_6^2 \,,
 }
where $d\tilde{s}_6^2$ is, for example, the metric of a Calabi-Yau three-fold, as in \cite{Candelas:1985en}.  This metric has the symmetries of Minkowski space, ${\bf R}^{3,1}$: translation invariance in the $t$ and $\vec{x} = (x^1,x^2,x^3)$ directions, and also rotation and boost invariance.

A large literature has grown up on the study of ``warped compactifications'' of string theory: for a review, see \cite{Douglas:2006es}.  A common way to write the metric ansatz is
 \eqn{Warped}{
  ds_{10}^2 = e^{2A(y)} \left[ -dt^2 + d\vec{x}^2 \right] + 
    e^{-2A(y)} d\tilde{s}_6^2 \,,
 }
where $y$ denotes, collectively, the coordinates involved in the metric $d\tilde{s}_6^2$.  This is usually thought to be the most general ansatz consistent with the symmetries of ${\bf R}^{3,1}$.

It is less than straightforward to construct solutions of string theory of the form \eno{Warped} where the extra-dimensional manifold is compact, because of a constraint from the strong energy condition \cite{Gibbons:1984kp,Maldacena:2000mw}:\footnote{If we define $\tilde{T}_{MN} = T_{MN} - {1 \over d-2} g_{MN} T^L_L$ in a $d$-dimensional theory, then the strong energy condition says that $\tilde{T}_{MN} \xi^M \xi^N \geq 0$ when $\xi^M$ is timelike or null.  Assuming the Einstein equations, $R_{MN} = \kappa_d^2 \tilde{T}_{MN}$, hold, this means that $R_{MN} \xi^M \xi^N \geq 0$.  In particular, $R^{00} \geq 0$ when the metric is diagonal.}
 \eqn{StrongCondition}{
  R^{00} = \tilde\square A \geq 0 \,,
 }
where $\tilde\square$ is the laplacian built from the metric $d\tilde{s}_6^2$.  The problem is that the integral of $\tilde\square A$ over the compact manifold vanishes, which can only happen if the inequality is saturated everywhere, meaning that the function is harmonic.  And a harmonic function on a compact manifold must be constant.  A resolution is to use orientifold planes: for example, O3-planes, as in \cite{Verlinde:1999fy}.  Near an O3-plane, the geometry is ill-defined: formally, $e^{-4A}$ becomes negative.

One could consider a more general ansatz:
 \eqn{TimeWarp}{
  ds_{10}^2 = e^{2A(y)} \left[ -h(y) dt^2 + d\vec{x}^2 \right] + ds_6^2 \,.
 }
This has all the same symmetries as before, except for boost invariance.  Boost invariance is quite well established experimentally (for a review, see for example \cite{Mattingly:2005re}), so one might dismiss the ansatz \eno{TimeWarp} as obviously unacceptable.  But suppose, for some reason, we are constrained to live at a particular value of $y$, call it $y_*$; or perhaps we are restricted to some narrow range of values of $y$ close to $y_*$ where $h$ is nearly constant.  Then we would perceive the world to be boost-invariant (or nearly so), with a speed of light $c = \sqrt{h(y_*)}$.  If $h(y) > h(y_*)$ away from $y_*$, then a particle that can propagate into the extra dimensions can appear to move superluminally from the point of view of an observer at $y_*$, in the sense that in a coordinate time $\Delta t$, the particle could propagate significantly further than $\sqrt{h(y_*)} \Delta t$ without violating causality in the extra-dimensional geometry.

The ideas in the previous paragraph have a long history, which I will trace only partially here.  Metrics of approximately the form \eno{TimeWarp} were discussed as early as \cite{Kaelbermann:1998hu}, with a qualitative hint arising even in \cite{Rubakov:1983bb}; and the special case $A=0$ was treated in \cite{Visser:1985qm}.  Time dependent versions were studied in \cite{Chung:1999xg,Youm:2001sw} in an effort to use a variable speed of light to solve cosmological problems without inflation, along the lines of \cite{Moffat:1992ud,Albrecht:1998ir}.  Further work in the direction of a variable speed of light has been reviewed in \cite{Magueijo:2003gj}, and the extensive topic of brane world cosmology has been reviewed in \cite{Langlois:2002bb}.  Five-dimensional metrics similar to \eno{TimeWarp} were termed ``asymmetrically warped spacetimes'' in \cite{Csaki:2000dm}, where, in the spirit of earlier work \cite{Randall:1999vf,Kiritsis:1999tx,Alexander:1999cb,Bowcock:2000cq}, the examples of five-dimensional AdS-Schwarzschild and Reissner-Nordstr\"om-AdS were discussed.  It was noted in \cite{Csaki:2000dm} that one needs $w < -1$ on the Planck brane.  The need for a violation of the null energy condition was later demonstrated more generally \cite{Cline:2001yt}, using an argument which I extend in this paper.  Special cases of \eno{TimeWarp} where $A=0$ were discussed in \cite{Dubovsky:2001fj}, and also in \cite{Deffayet:2001aw} as part of an approach to the cosmological constant problem using the model of \cite{Dvali:2000hr}.  Some general constraints on asymmetrically warped string theory constructions were considered in \cite{Frey:2003jq}.  More explicit string theory constructions have been considered, for example in \cite{Ganor:2006ub}.

Returning to the ansatz \eno{TimeWarp}: It's difficult to arrange for non-constant $h(y)$ over a compact extra-dimensional manifold because of a constraint arising from the null energy condition:\footnote{The null energy condition says $T_{MN} \xi^M \xi^N \geq 0$ for all null vectors $\xi^M$.  Assuming the Einstein equations hold, this means $R_{MN} \xi^M \xi^N \geq 0$.  In particular, $-R^0_0+R^1_1 \geq 0$ when the metric is diagonal.}
 \eqn{NullCondition}{
  4h^2 e^{-2A} (-R^0_0 + R^1_1) = 
   -3\tilde{g}^{mn} \partial_m h \partial_n h + 
     \tilde\square (h^2) \geq 0 \,.
 }
Integrating over the compact manifold (and supposing $e^{-2A}$ is for some reason well-defined everywhere), one would be forced by the inequality to conclude that $h$ is constant.  This argument is in the same spirit as \cite{Cline:2001yt}, but it generalizes more easily to (almost) any dimension, as we will see in section~\ref{BULK}.  In contrast to the situation described by \eno{StrongCondition}, I do not know an explicit string theory construction that would evade the no-go argument based on \eno{NullCondition}.  However, one may temporarily ignore this argument by considering non-compact extra dimensions.  This amounts to turning off gravity, because the wave-function of the four-dimensional graviton is non-normalizable in the extra dimensions.

The simplest sort of non-compact asymmetrically warped geometry is just an extra-dimensional black brane.  For example, the near-extremal D3-brane has a metric of the form \eno{TimeWarp} with
 \eqn{Dthree}{
  d\tilde{s}_6^2 = e^{-2A(y)} \left( 
    {dy^2 \over h(y)} + y^2 d\Omega_5^2 \right) \,,
 }
where now $y$ is a single real variable, $d\Omega_5^2$ is the metric on a unit $S^5$, and
 \eqn{FoundH}{
  e^{-4A(y)} = 1 + {L^4 \over y^4} \qquad\qquad
   h(y) = 1 - {y_0^4 \over y^4} \,.
 }
A difficulty is that the existence of a regular horizon at $y=y_0$ is associated with a finite Hawking temperature,
 \eqn{HawkingT}{
  T = {1 \over \pi y_0} \left( 1 + {L^4 \over y_0^4} \right)^{-1/2}
   \,.
 }
Following \cite{Kiritsis:1999tx}, we might imagine our world as a brane at some fixed value of $y$ in the geometry \eno{Dthree}.  Certainly, such a construction would lead to an observed speed of light which is slower than what can be attained far from the D3-branes.  But the finite temperature \eno{HawkingT} makes the construction seem less interesting, because it means we are not describing the ground state.  Likewise, as emphasized in \cite{Creminelli:2001tc}, the $AdS_5$-Schwarzschild and $AdS_5$-Reissner-Nordstr\"om geometries considered in \cite{Csaki:2000dm} do not describe ground states, but instead finite temperature states of a strongly coupled conformal theory.  An exception is the extremal $AdS_5$-Reissner-Nordstr\"om solution, which has zero temperature; but it still has a macroscopic Bekenstein-Hawking entropy, meaning that it doesn't describe a single physical state, but instead a large ensemble of states.  This may not be fatal, but entropic solutions do not seem to me the best of starting points when seeking to describe the vacuum.

I wish to consider in this paper a subclass of asymmetrically warped solutions which, by requirement, have no temperature and no entropy.  I will refer to such backgrounds as ``time warps.''  A simple way to construct one is to cut off a black brane background above the horizon, as was indeed considered in \cite{Csaki:2000dm,Cline:2003xy}.  I will instead work with five-dimensional variants of the geometries found in \cite{Gubser:2008wz} in asymptotically $AdS_4$ geometries, using the abelian Higgs model coupled to gravity.  This model, first considered in \cite{Gubser:2008px}, has the following action:
 \eqn{Action}{
  S_{\rm bulk} = {1 \over 16\pi G_{D+1}} \int d^{D+1} x \, \sqrt{g} 
    {\cal L}_{\rm bulk} \,,
 }
where
 \eqn{Lagrangian}{
  {\cal L}_{\rm bulk} = R - {1 \over 4} F_{\mu\nu}^2 - 
    |(\partial_\mu - i q A_\mu) \psi|^2 - V(|\psi|) \,.
 }
For appropriate choices of $V(|\psi|)$ and $q$, the classical equations of motion following from \eno{Action} admit superconducting black hole solutions \cite{Gubser:2008px,Hartnoll:2008vx,Hartnoll:2008kx,Gubser:2008pf}, which spontaneously break the $U(1)$ gauge symmetry in the bulk.  See \cite{Gubser:2005ih} for earlier work on superconducting black holes; \cite{Hartnoll:2007ih,Hartnoll:2007ip} for earlier work on the possible relation between black holes in $AdS_4$ and phases of superconducting materials; \cite{Basu:2008st,Herzog:2008he}, among others, for discussion of variants on this type of solution and dual superfluids; and \cite{Sachdev:2008ba} for an overview and further references.

The focus on four-dimensional anti-de Sitter space in the superconducting black hole literature has been driven by the interest in strongly coupled $2+1$-dimensional conformal field theories, which has been driven in turn by the hope of explaining phenomena in thin-film or layered superconductors in terms of quantum critical points.\footnote{An exception is \cite{Horowitz:2008bn}, which deals with five-dimensional geometries.  Also there have been studies of superconducting black holes based on the Einstein-Yang-Mills lagrangian in four \cite{Gubser:2008zu,Gubser:2008wv,Roberts:2008ns} and higher \cite{Manvelyan:2008sv} dimensions.}  In this paper I will focus instead on five-dimensional geometries, because they provide a minimal example of non-compact time warp geometries that include a copy of ${\bf R}^{3,1}$.  By minimal, I mean that there is only one extra dimension, and the field content appears to be the minimal one that can support a time warp.  I will find the geometries using the same strategy as in \cite{Gubser:2008wz}.  And as in \cite{Gubser:2008wz}, the time warp geometries I find are domain walls with conformal invariance in the ultraviolet (UV) and infrared (IR) regions of the bulk, but different speeds of light as measured by a fixed coordinate system on the boundary.  One therefore expects that correlation functions interpolate between a conformally invariant form in the ultraviolet, and a different conformally invariant form---characterized by a different speed of light---in the infrared.

The rest of this paper is organized as follows.  In section~\ref{BULK} I exhibit the time warp geometry which I focus on.  In section~\ref{CORRELATORS} I calculate the spectral measure of two-point correlators for a massive scalar propagating in the time warp geometry.  I relate this spectral measure to unparticle phase space and explain how time warp effects could manifest themselves in the decay of a heavy particle into a light visible particle plus unparticle stuff.  An appendix contains some additional technical detail on the computation of two-point functions.  In section~\ref{BRANE}, I consider what it would take to ``compactify'' the asymptotically $AdS_5$ time warp geometries using a Planck brane construction.\footnote{``Compactify'' is a slight misnomer here because after introducing the Planck brane, the geometry is still non-compact in the infrared, as in \cite{Randall:1999vf}.  It is like a true compactification in that there is a finite five-dimensional volume below any finite four-volume element on the Planck brane.}  As one might anticipate based on the no-go argument discussed around \eno{NullCondition}, I am forced to entertain a theory on the brane which violates the null energy condition.  In section~\ref{GRAVITON}, I show that it is possible to obtain an four-dimensional infrared-massless graviton by including a wrong-sign Einstein-Hilbert term in the action of the Planck brane.  Altogether, it must be admitted that the Planck brane construction is peculiar.

\section{Solutions in the bulk}
\label{BULK}

I will construct a solution to the classical equations of motion of the Abelian Higgs model in $AdS_5$, starting with the ansatz
 \eqn{AdSansatz}{
  ds_5^2 = e^{2A(r)} \left[ -h(r) dt^2 + d\vec{x}^2 \right] + 
    e^{2B(r)} dr^2 \,.
 }
$B(r)$ parametrizes the gauge freedom of choosing different radial variables.  Let us pass to the gauge $B=-{1 \over 2} \log h$, where the equations of motion and constraints are simplest to state.  Rotational invariance in the $\vec{x}$ directions forces $A_i=0$ for $i=1,2,3$.  Let's also assume $A_r=0$ (which is a gauge choice) and use $\Phi$ to denote $A_0$.  Finally, let's restrict attention to solutions where $\psi$ is real.  The equations of motion are
 \begin{subequations}\label{eomsZero}
 \begin{eqnarray}
  A'' &=& -{1 \over 3} |\psi'|^2 - 
    {e^{-2A} \over 3h^2} q^2 \Phi^2 |\psi|^2  \label{eomsZeroA} \\
  h'' + 4 A' h' &=& e^{-2A} \left( \Phi'^2 + 
    {2 \over h} q^2 \Phi^2 |\psi|^2 \right)  \label{eomsZeroB} \\
  \Phi'' + 2 A' \Phi' &=& {2 \over h} q^2 \Phi |\psi|^2  \label{eomsZeroC} \\
  \psi'' + \left( 4A' + {h' \over h} \right) \psi' &=& 
    -{e^{-2A} \over h^2} q^2 \Phi^2 \psi + 
    {1 \over h} {\partial V \over \partial\psi^*} \,. \label{eomsZeroD} 
 \end{eqnarray}
 \end{subequations}
The constraint coming from the $G_{rr}$ Einstein equation is
 \eqn{constraint}{
  24 h^2 A'^2 + 6 hh' A' + e^{-2A} h\Phi'^2 - 2h^2 |\psi'|^2 - 
   2e^{-2A} q^2 \Phi^2 |\psi|^2 + 2 h V = 0 \,.
 }
An additional first-order equation can be extracted by noticing that the quantity
 \eqn{NoetherCharge}{
  Q = e^{4A} (h' - e^{-2A} \Phi \Phi')
 }
is a constant if the equations \eno{eomsZero} and \eno{constraint} are obeyed.  Assuming that the infrared geometry is asymptotically $AdS_5$ amounts to assuming that $Q=0$.  This is because both $h'$ and $\Phi'$ must go to zero in the infrared, while $h$ remains non-zero and $\Phi$ is bounded.  Solutions with a regular horizon typically have $Q \neq 0$.  So $Q=0$ is a sort of extremality condition.\footnote{The existence of the conserved charge \eno{NoetherCharge} and its relevance to extremal solutions were pointed out to me by A.~Nellore.  In a more general gauge where $B$ is left as an unspecified function of $r$ but $A_r$ is still constrained to vanish,
 \eqn{NoetherChargeGeneral}{
  Q = {e^{4A-B} \over \sqrt{h}} (h' - e^{-2A} \Phi \Phi') \,.
 }
It may seem puzzling that $Q$ isn't invariant under shifts of $\Phi$ by a constant, given that such shifts preserve the more general gauge just described.  However, after such a shift, the complex scalar $\psi$ would have a time-dependent phase.  Requiring $\psi$ to be time-independent thus disallows shifts of $\Phi$ by a constant---unless $\psi=0$.  If in fact $\psi=0$, then there is an additional conserved quantity, ${e^{2A-B} \over \sqrt{h}} \Phi'$.  Shifting $\Phi$ by an additive constant causes $Q$ to change by some multiple of this quantity.}

Already from \eno{eomsZeroA} and \eno{eomsZeroB} one can see the no-go theorem of \cite{Cline:2001yt} at work, as well as a application of the $c$-theorem argument of \cite{Freedman:1999gp}.  In brief: the left-hand side of \eno{eomsZeroA} is proportional to $R^0_0 - R^r_r$, so the right hand side has to be negative according to the null energy condition.  Thus $A$ is superharmonic as a function of $r$, which means it can't have a minimum unless one introduces an additional matter source.  The left-hand side of \eno{eomsZeroB} is proportional to $R^1_1 - R^0_0$, so the right hand side has to be positive, again according to the null energy condition.  Thus $h$ is subharmonic with respect to the line element $d\tilde{s}^2 = e^{-8A} dr^2$ in the $r$ direction, indicating that it cannot have a maximum without some additional matter source as long as $A$ is well defined.  As argued in \cite{Gubser:2008wz}, \eno{eomsZeroA} and \eno{eomsZeroB} also imply that for backgrounds which are asymptotically anti-de Sitter, the effective speed of light $\sqrt{h(r)}$ increases from the infrared to the ultraviolet.\footnote{There appears to be some tension between \eno{StrongCondition}, which indicates that $A$ is subharmonic in a ten-dimensional compactification, and \eno{eomsZeroA}, which shows that it is superharmonic in five.  To understand the situation more comprehensively, consider a $D$-dimensional ansatz
 \eqn{DDimension}{
  ds_D^2 = e^{2A} (-h dt^2 + d\vec{x}^2) + e^{2B} d\tilde{s}_{D-4}^2
    \,,
 }
where $A$, $B$, and $h$ depend only on the $D-4$ coordinates of $d\tilde{s}^2_{D-4}$.  Setting $h=1$, $D \neq 6$, and $B = {4 \over 6-D} A$, one finds
 \eqn{StrongD}{
  -e^{2B} R^0_0 = \tilde\square A \,,
 }
and the strong energy condition says this should be positive.  Both the strong energy condition and the null energy condition are satisfied by the stress tensor following from \eno{Lagrangian}.  Indeed, if $h=1$ and $D=5$, $A$ is superharmonic with respect to the metric $dr^2$ and subharmonic with respect to the metric $e^{8A} dr^2$.

Using the same ansatz \eno{DDimension}, assuming $D \neq 6$, and setting $B = {4 \over 6-D} A$, one finds that
 \eqn{NullAgain}{
  4h^2 e^{2B} (-R^0_0 + R^1_1) = -3 \tilde{g}^{mn} \partial_m h
    \partial_n h + \tilde\square (h^2) \,.
 }
So the argument around \eno{NullCondition} survives essentially unchanged for dimensions $D \neq 6$.

It is notable that both lines of argument discussed in this footnote have no force for $D=6$.  It would be interesting to consider six-dimensional time warps in more detail.}
  
Unfortunately, I don't know a choice of the scalar potential that leads to analytically tractable equations of motion, even in the presence of the extremality constraint $Q=0$.  So I will solve them numerically for particular choices of parameters.  Specifically, I will choose
 \eqn{VChoice}{
  V = -{12 \over L^2} + m^2 |\psi|^2 + {u \over 2} |\psi|^4 \,,
 }
with $m^2 < 0$ and $u>0$.  With this choice, the infrared geometry is itself anti-de Sitter, but with a different radius, $L_{\rm IR}$, determined through the equation $V(\psi_{\rm IR}) = -12/L_{\rm IR}^2$, where $\psi_{\rm IR} = \sqrt{-m^2/u}$ is the $U(1)$ symmetry-breaking extremum of the potential.  The infrared copy of $AdS_5$ signals emergent conformal symmetry: emergent in the sense that it arises only in the infrared limit of the dual field theory.  It was speculated in \cite{Gubser:2008wz} that scalar potentials which have no minima lead to emergent Lorentz symmetry in the infrared, rather than emergent conformal symmetry.  Evidence in favor of this conjecture has appeared in \cite{Gubser:2008pf}.

For numerical work, I found that the most convenient gauge is $B=0$, instead of the gauge $B = -{1 \over 2} \log h$ that I used in \eno{eomsZero}-\eno{NoetherCharge}.  In the $B=0$ gauge, $A''$ does not have a definite sign.  However, $A' = 1/L$ in the UV and $A' = 1/L_{\rm IR}$ in the IR\@.

 \begin{figure}
  \centerline{\includegraphics[width=4.5in]{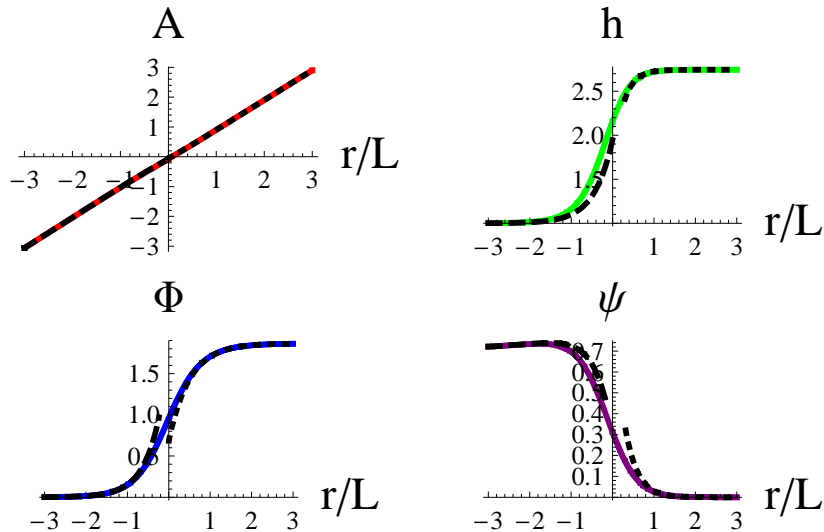}}
  \caption{(Color online.)  A time warp geometry for the choice of parameters~\eno{ExampleParameters}, in the gauge $B=0$.  Solid colored curves are from numerics; dashed black curves are infrared asymptotics; and dotted black curves are ultraviolet asymptotics.}\label{EXAMPLE}
 \end{figure}
The ultraviolet dimension of the operator dual to $\psi$ is
 \eqn{DualDim}{
  \Delta_\psi^{\rm UV} = 2 + \sqrt{4 + m^2 L^2} \,.
 }
Provided $-4 \leq m^2 L^2$, the asymptotically $AdS_5$ geometry is stable \cite{Breitenlohner:1982bm,Breitenlohner:1982jf}.  If also $m^2 L^2 \leq 0$, then one does not have to stipulate boundary conditions on $\psi$ at the conformal boundary.  However, one may do so, and in studying solutions to \eno{eomsZero}, I generally did: I required $\psi \propto e^{-\Delta_\psi^{\rm UV} A}$ rather than $\psi \propto e^{(\Delta_\psi^{\rm UV}-4)A}$.  My choice corresponds to requiring that the breaking of the $U(1)$ symmetry associated with the phase of $\psi$ is spontaneous rather than explicit in the dual conformal theory.\footnote{It is well understood \cite{Klebanov:1999tb} that for the range $-4 \leq m^2 L^2 < -3$, one can replace $\Delta_\psi^{\rm UV} \to 4-\Delta_\psi^{\rm UV}$; this corresponds to a new CFT, from which the original can be recovered from a renormalization group flow triggered by double-trace terms.  It is even possible to make sense of more general boundary conditions on $\psi$ in terms of multi-trace operators \cite{Witten:2001ua}.}  With this choice (or with any definite choice of boundary conditions on $\psi$), according to the same reasoning as in \cite{Gubser:2008wz}, there can be at most discretely many solutions with $h$ nowhere vanishing.  In instances where I was able to find more than one solution, the one with no nodes in $\psi$ had the smallest value of $h$ in the UV.\footnote{For example, the choice of parameters \eno{ExampleParameters} leads to one solution with $h_{\rm UV} \approx 2.74$, which I will describe in some detail, and another with $h_{\rm UV} \approx 34.0$, in which $\psi$ has a single node.}  I will assume that solutions with no nodes (or as few nodes as possible) in $\psi$ are preferred; however, this is just an assumption.  In figure~\ref{EXAMPLE}, I exhibit the solution I found for the following choice of parameters:
 \eqn{ExampleParameters}{
  qL = 3 \qquad m^2 L^2 = -2 \qquad uL^2 = 4 \,.
 }
In the solution I found for this choice of parameters, the speed of light is about $1.7$ times faster in the ultraviolet part of the geometry (large positive $r$) than in the infrared part (large negative $r$).  With the choice of parameters \eno{ExampleParameters}, it happens that $L_{\rm IR}$ and $L$ are quite close together: $L_{\rm IR}/L \approx 0.97$.

It is possible to generate series expansions in the infrared and the ultraviolet for the solution I have described.  In the infrared,
 \begin{subequations}\label{IRasymp}
 \begin{eqnarray}
  A &=& {r \over L_{\rm IR}} + \ldots \label{IRasympA}  \\
  h &=& 1 + 
    {\Delta^{\rm IR}_\Phi-2 \over \Delta^{\rm IR}_\Phi-3} \Phi_1^2
     e^{2(\Delta_\Phi^{\rm IR}-3) r/L_{\rm IR}} + \ldots
     \label{IRasymph}  \\
  \Phi &=& \Phi_1 e^{(\Delta_\Phi^{\rm IR}-2)r/L_{\rm IR}} + 
     \ldots \label{IRasympPhi}  \\
  \psi &=& \psi_{\rm IR} + 
     \psi_1 e^{(\Delta_\psi^{\rm IR}-4)r/L_{\rm IR}} + 
     \ldots \label{IRasymppsi} \,,
 \end{eqnarray}
 \end{subequations}
where $\ldots$ denotes terms that are exponentially smaller in the infrared than the ones shown.  Here
 \eqn{DeltaIRdefs}{
  \Delta_\Phi^{\rm IR} &= 1 + \sqrt{1 + 2 q^2 
    \psi_{\rm IR}^2} L_{\rm IR}^2  \cr
  \Delta_\psi^{\rm IR} &= 2 + \sqrt{4 + 
    (m^2 + 3u\psi_{\rm IR}^2) L_{\rm IR}^2} \,,
 }
and the solution I exhibited has
 \eqn{FoundIRcoefs}{
  \Phi_1 = 1 \qquad \psi_1 = 0.168 \,.
 }
(A scaling symmetry, $x^m \to \lambda x^m$ while $r \to r - L_{\rm IR} \log\lambda$ and $\Phi \to \Phi/\lambda$, allows one to set $\Phi_1=1$ provided it is non-zero.)  In the ultraviolet,
 \begin{subequations}\label{UVasymp}
 \begin{eqnarray}
  A &=& {r \over L} + a_1 - {p_1 p_2 \over 16 h_{\rm UV}}
    e^{-4r/L} + \ldots \label{UVasympA}  \\
  h &=& h_{\rm UV} + {p_1 p_2 \over 2} e^{-4r/L} + \ldots
     \label{UVasymph}  \\
  \Phi &=& p_1 + p_2 e^{-2r/L} + \ldots \label{UVasympPhi}  \\
  \psi &=& s_2 e^{-\Delta_\psi^{\rm UV} r/L} + 
     \ldots \label{UVasymppsi} \,,
 \end{eqnarray}
 \end{subequations}
where $\ldots$ indicates terms which are exponentially smaller in the ultraviolet than the ones shown.  The solution I exhibited has
 \eqn{FoundUVcoefs}{
  a_1 = -0.0963 \qquad
  h_{\rm UV} = 2.74 \qquad
  p_1 = 1.86 \qquad
  p_2 = -1.18 \qquad
  s_2 = 0.955 \,.
 }
The asymptotics shown in figure~\ref{EXAMPLE} are based on \eno{IRasymp} and~\eno{UVasymp}, but in some cases evaluated to higher orders and with greater numerical precision.

\section{Green's functions and unparticles}
\label{CORRELATORS}

If we think of the UV speed of light as the ordinary one of daily experience, then the asymptotically anti-de Sitter geometry can be interpreted as describing a medium with a definite index of refraction.  Low-energy signals pass through the infrared part of the geometry, where they can only go at a fraction of the UV speed of light.

A more tantalizing possibility is that the physics of our world can be described as the infrared physics in a time warp geometry.  Then the speed of light that we measure is the infrared speed of light.  We are, in effect, caught in the refractive medium.  If only we could pass through the domain wall and get into the UV part of the geometry, then we could move superluminally from the perspective of an infrared observer.

As a first step toward exploring the phenomenology of time warps, I consider in this section how relativistic kinematics is altered for the field theory dual to the type of non-compact time warp constructed in section~\ref{BULK}.  Because such a geometry is asymptotically $AdS_5$, one can extract two-point correlators for operators in the dual field theory.  This field theory is strongly coupled, so it is not straightforward to make a comparison with perturbative quantum field theory.  What is straightforward is to make a connection with recent ideas about ``unparticle physics'' \cite{Georgi:2007ek}, which is the possibility that an approximately conformal, strongly coupled sector---the unparticles---will be discovered through high-energy collisions, for example at the LHC\@.  In section~\ref{SPECTRAL}, without reference to time warps, I review the connection between the imaginary part of the two-point Green's functions in a conformal field theory and the phase space for unparticles.  Then, in section~\ref{PHASEMOD}, I calculate in a specific example how this phase space gets modified by time warp effects.

\subsection{Spectral measure and unparticle phase space}
\label{SPECTRAL}

First let's recall how multi-particle phase space measure is related to the imaginary part of an appropriate Green's function.  Let $\phi$ be a canonically normalized, free, massless, real scalar in four dimensions.  For any integer $n \geq 1$, the operator ${\cal O} = \phi^n$ has dimension $\Delta = n$.  Its time-ordered Green's function is
 \eqn{GF}{
  G_F(x) &\equiv -i \langle 0 | T \left\{ {\cal O}(x) {\cal O}(0)
    \right\} | 0 \rangle \,.
 }
It's an exercise in free field theory to verify that the phase space measure for $n$ outgoing $\phi$ particles, collectively carrying four-momentum $k_\mu = (-\omega,\vec{k})$, is
 \eqn{PhaseMeasure}{
  d\Phi(k) = -2\theta(\omega) \Im G_F(k) {d^4 k \over (2\pi)^4} \,.
 }
Because the right hand side is essentially the spectral measure for the Green's function $G_F$, it makes sense to use precisely the same expression for ``unparticle phase space'' even when ${\cal O}$ has no construction in terms of free fields.  In such a case, the only constraint on the dimension $\Delta$ is the unitarity bound $\Delta \geq 1$.  Using \eno{PhaseMeasure} for general conformal field theories is part of the proposal of \cite{Georgi:2007ek}, although it was phrased a little differently there.

Let's now review the computation of $\Im G_F(k)$ from a dual geometry which is asymptotically $AdS_5$ with radius $L$.  I will assume that ${\cal O}$ is dual to a real, minimally coupled scalar $\phi$ in five dimensions.  To quadratic order, its lagrangian is
 \eqn{Lphi}{
  {\cal L}_\phi = -{1 \over 2} (\partial\phi)^2 - 
    {1 \over 2} m_\phi^2 \phi^2 \,,
 }
up to a prefactor which I will not try to track.  The computation of imaginary parts of real-time two-point Green's functions is familiar from literature (for example \cite{Das:1996wn,Klebanov:1997kc,Gubser:1997cm,Gubser:1997se}) predating AdS/CFT \cite{Maldacena:1997re,Gubser:1998bc,Witten:1998qj}: it hinges on the identification of a conserved flux.  Related computations were revisited in \cite{Son:2002sd} and shown to follow from the original AdS/CFT prescription upon appropriate use of Schwinger-Keldysh contours \cite{Herzog:2002pc}.  See \cite{Gubser:2008yx,Gubser:2008sz} for a recent application to the computation of bulk viscosity which is fairly similar to the calculation of interest here.  The standard procedure is to find solutions to the equations of motion following from \eno{Lphi} of the form
 \eqn{phiAnsatz}{
  \phi(t,\vec{x},r) = e^{-i\omega t + i\vec{k} \cdot \vec{x}} 
    f_k(r) \,,
 }
where $f_k(r)$ is required to satisfy appropriate boundary conditions in the infrared---to be discussed below.  In general, the functions $f_k(r)$ are complex.  When they are, $f_k^*(r)$ satisfies the same radial equation that $f_k(r)$ does, because all the coefficients in the radial equation are real functions of $r$.  As a consequence of Abel's identity, the flux
 \eqn{FluxDef}{
  {\cal F}_k \equiv L h e^{4A-B} \Im f_k^* \partial_r f_k
 }
is conserved (meaning independent of $r$).  The imaginary part of the two-point function of ${\cal O}$ is then evaluated as
 \eqn{GpEval}{
  \Im G_F(k) = \lim_{r \to \infty} 
    K_{\cal O} \left( {L \over 2} \right)^{4-2\Delta} 
     e^{2(\Delta-4)A} {{\cal F}_k \over |f_k(r)|^2} \,,
 }
where $K_{\cal O}$ is a positive, dimensionless prefactor related to how the lagrangian \eno{Lphi} is normalized.

In pure $AdS_5$, where $h=1$ and $A = r/L$, one straightforwardly finds
 \eqn{AdSGF}{
  f_k(r) = \left\{ \seqalign{\span\TR & \qquad\span\TT}{
    e^{-2r/L} H_{\Delta-2}^{(1)}(L \sqrt{\omega^2-\vec{k}^2}
      e^{-r/L}) & for $\omega^2 > \vec{k}^2$  \cr
    e^{-2r/L} K_{\Delta-2}(L \sqrt{\vec{k}^2-\omega^2}
      e^{-r/L}) & for $\omega^2 < \vec{k}^2$ \,.
   } \right.
 }
These solutions satisfy the infrared boundary conditions appropriate for computing the time-ordered propagator $G_F$: infalling when $\omega > |\vec{k}|$; outgoing when $\omega < -|\vec{k}|$; and decaying rather than growing in the infrared when $\omega^2 < \vec{k}^2$.  Plugging \eno{AdSGF} into \eno{GpEval}, one obtains
 \eqn{AdSResult}{
  \Im G_F(k) = -{2\pi K_{\cal O} \over \Gamma(\Delta-2)^2}
    (\omega^2-\vec{k}^2)^{\Delta-2} \theta(\omega^2-\vec{k}^2) \,.
 }

\subsection{Time warp modification of unparticle phase space}
\label{PHASEMOD}

The relation \eno{PhaseMeasure} equates phase space with spectral measure.  The same relation can be used in the context of a time warp.  The only difference is that the spectral measure will no longer be capable of expression solely in terms of $-\omega^2 + \vec{k}^2$; instead, it is a function of $\omega$ and $|\vec{k}|$ separately.  The purpose of the current section is to examine the spectral measure and explain how it affects a decay process that involves unparticles.

In a time warp geometry which is asymptotically $AdS_5$ in both the UV and IR, one expects the following asymptotic forms for small and large $\omega$, respectively:
 \eqn{LimitingForms}{
  \Im G_F^{\rm IR}(k) &= -{2\pi K_{\cal O}^{\rm IR} \over
    \Gamma(\Delta_\phi^{\rm IR}-2)^2} 
    (\omega^2-\vec{k}^2)^{\Delta_\phi^{\rm IR}-2}
    \theta(\omega^2-\vec{k}^2)  \cr
  \Im G_F^{\rm UV}(k) &= -{2\pi K_{\cal O}^{\rm UV} \over
    \Gamma(\Delta_\phi^{\rm UV}-2)^2} 
    (\omega^2/h_{\rm UV}-\vec{k}^2)^{\Delta_\phi^{\rm UV}-2}
    \theta(\omega^2/h_{\rm UV}-\vec{k}^2) \,.
 }
The UV dimension $\Delta_\phi^{\rm UV}$ is given simply by the formula
 \eqn{DeltaPhiUV}{
  \Delta_\phi^{\rm UV} = 2 + \sqrt{4 + m_\phi^2 L^2} \,.
 }
To calculate the IR dimension $\Delta_\phi^{\rm IR}$ one must replace $L$ by the radius $L_{\rm IR}$ of the infrared copy of $AdS_5$.  The dimensionless parameter $K_{\cal O}^{\rm UV}$ is related to how the lagrangian \eno{Lphi} is normalized, just as $K_{\cal O}$ was in the discussion above.  $K_{\cal O}^{\rm IR}$ is a dimensionless multiple of $K_{\cal O}^{\rm UV} L^{2(\Delta_\phi^{\rm IR}-\Delta_\phi^{\rm UV})}$, and it can be calculated once the background geometry is known.

Note that the condition for a momentum to be UV-timelike is $\omega^2/h_{\rm UV}-\vec{k}^2 > 0$.  Because $h_{\rm UV}>1$, a UV-timelike momentum is necessarily IR-timelike; but an IR-timelike momentum may be UV-timelike or UV-spacelike.  This is because the momentum $k_\mu = (-\omega,\vec{k})$ is a covariant vector, i.e.~a $1$-form.  Thus $(k^2)_{\rm UV} = -\omega^2/h_{\rm UV} + \vec{k}^2$, while $(k^2)_{\rm IR} = -\omega^2 + \vec{k}^2$.  The opposite conclusion would be reached for contravariant vectors like an infinitesimal displacement $dx^\mu$: if it is UV-timelike, it can be either IR-timelike or IR-spacelike.  The latter possibility is what one would mean by an infinitesimal faster-than-light displacement.  It makes sense that the spectral measure of Green's functions occupies a narrower light-cone in momentum space in the UV limit than in the IR, because causal trajectories in the UV limit occupy a broader light-cone in real space.  The retarded Green's function $G_R(t,\vec{x})$ must be non-zero over the broader position-space light-cone defined by the UV speed of light, but at large separations, I expect that $G_R$ is very attenuated outside the narrower position-space light-cone defined by the IR speed of light.  Colloquially: You can go faster than light, but perhaps not for long.

To examine how the Green's function interpolates between the limiting behaviors shown in \eno{LimitingForms}, one may consider a dimensionless phase space modification factor:
 \eqn{WarpRatio}{
  W(k) \equiv {\Im G_F(k) \over \Im G_F^{\rm IR}(k)} \,.
 }
$W(k)$ is to be evaluated only for infrared-timelike momenta.  One should find $W(k) \to 1$ as $\omega \to 0$.  For large $\omega$, according to \eno{LimitingForms}, one should find
 \eqn{WarpRatioUV}{
  W(k) \propto {(\omega^2/h_{\rm UV}-\vec{k}^2)^{\Delta_\phi^{\rm UV}
    -2} \over (\omega^2-\vec{k}^2)^{\Delta_\phi^{\rm IR}-2}}
   \theta(\omega^2/h_{\rm UV}-\vec{k}^2) \,.
 }
 \begin{figure}
  \centerline{\includegraphics[width=4in]{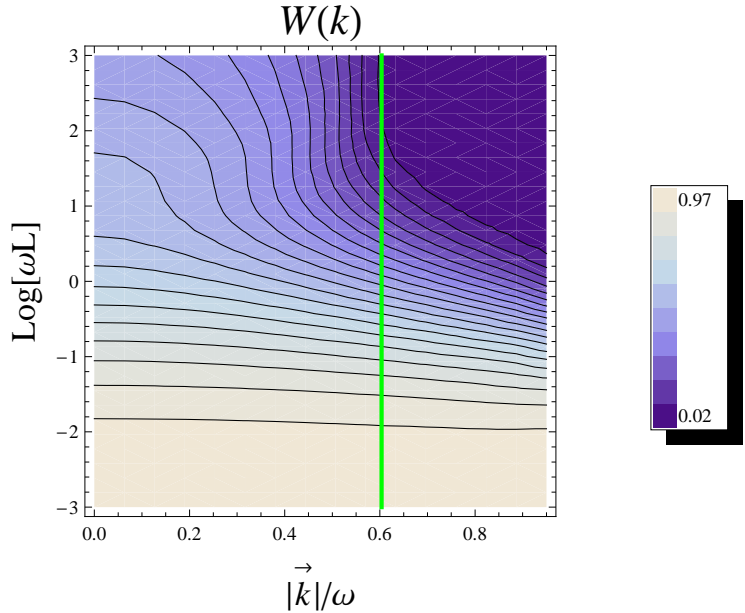}}
  \caption{(Color online.)  The phase space ratio $W(k)$ defined in \eno{WarpRatio}, for values of parameters discussed in the main text.  The vertical green line shows where UV-null momenta lie.}\label{WMOD}
 \end{figure} 
In figure~\ref{WMOD}, I show some numerical evaluations of $W(k)$ for the time warp geometry exhibited in figure~\ref{EXAMPLE} and for $m^2 L^2 = -\sqrt{10}$, corresponding to a dual operator with UV dimension $\Delta_\phi^{\rm UV} \approx 2.92$ and IR dimension $\Delta_\phi^{\rm IR} \approx 2.98$.  All indications from this figure, as well as further numerical studies, are that $W(k)$ does interpolate smoothly between $1$ in the IR and \eno{WarpRatioUV} in the UV\@.  In appendix~\ref{GREENS} I give some further details on the computation of two-point functions.

\begin{figure}
  \includegraphics[width=6in]{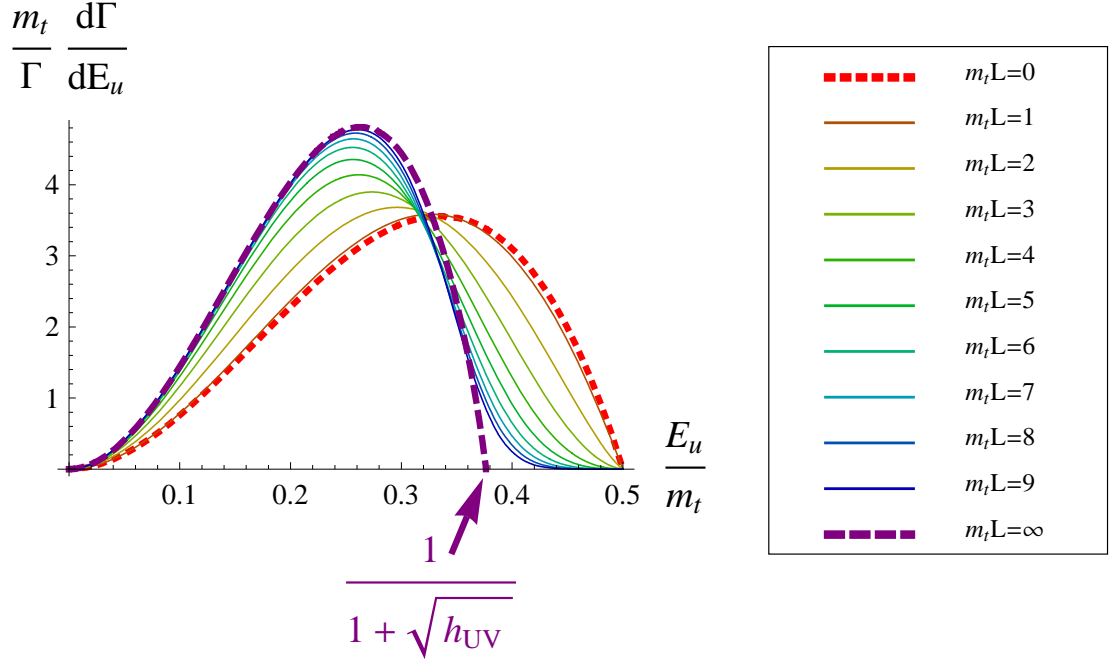}
  \caption{(Color online.)  The distribution of energies $E_u$ for the $u$ quark in the process $t \to u + {\cal U}$, where the unparticle stuff has infrared dimension $\Delta_\phi^{\rm IR} = 2.98$ and the time warp modifications are from the factor $W(k)$ plotted in figure~\ref{WMOD}.  Different curves come from different choices of the dimensionless parameter $m_t L$.  Each curve is normalized to have unit area under it.}\label{PHASECUTOFF}
 \end{figure}

With the modification factor $W(k)$ in hand, we can reconsider the process $t \to u + {\cal U}$ analyzed in \cite{Georgi:2007ek}.\footnote{There is no special reason to consider top and up quarks: any decay of a heavy visible-sector particle to a light particle plus unparticle stuff would serve as well.}  In order to make an explicit analysis, I use the $W(k)$ shown in figure~\ref{WMOD}, as well as the specific value $\Delta_\phi^{\rm IR} \approx 2.98$.  Also, I assume that all violations of infrared Lorentz invariance arise from $W(k)$.  That is, I assume that the $u$ quark propagates at the infrared speed of light, no matter what its energy; and I assume that the relevant coupling can be written as 
 \eqn{DecayCoupling}{
  {\cal L}_{\rm int} = i {\lambda \over 
    \Lambda^{\Delta_\phi^{\rm IR}}} \bar{u} \gamma_\mu
    (1-\gamma_5) t \, \partial^\mu {\cal O} + 
    \hbox{h.c.} \,,
 }
where $\Lambda$ is some high scale related to the mass of messenger fields that are integrated out to obtain \eno{DecayCoupling}.  The differential decay rate, expressed as a positive measure on phase space, is
 \eqn{DecayRate}{
  d\Gamma = {\overline{|{\cal M}|^2} \over 2m_t}
   (2\pi)^4 \delta^4(k_t - k_u - k_{\cal U})
    d\Phi_u(k_u) d\Phi_{\cal U}(k_{\cal U}) \,,
 }
where
 \eqn{PhaseSpaces}{
  d\Phi_u(k_u) &= \theta(\omega_u) 2\pi \delta(k_u^2)  \cr
  d\Phi_{\cal U}(k_{\cal U}) &=
    A_{\cal U} \theta(\omega_{\cal U}) 
      \theta(\omega_{\cal U}^2 - \vec{k}_{\cal U}^2)
      (\omega_{\cal U}^2 -
        \vec{k}_{\cal U}^2)^{\Delta_\phi^{\rm IR} - 2}
      W(k_{\cal U}) \,.
 }
The whole setup is just as in \cite{Georgi:2007ek} except for the factor $W(k_{\cal U})$ in \eno{PhaseSpaces}.  To obtain the distribution of up quark energies, we evaluate
 \eqn{UpEnergies}{
  {m_t \over \Gamma} {d\Gamma \over dE_u} \equiv
    {m_t \over \Gamma} \int d\Gamma \, \delta(E_u - \omega_u)
   \propto {E_u^2 \over m_t^2} \left[ 1 - {2E_u \over m_t}
     \right]^{\Delta_\phi^{\rm IR}-2} 
     \theta\left( {m_t \over 2} - E_u \right)
     W(k_{\cal U}) \,,
 }
where to find the last expression one must note that $\overline{|{\cal M}|^2} \propto E_u$.  If $m_t \gg 1/L$, then we can combine \eno{WarpRatioUV} and \eno{UpEnergies} to get
 \eqn{UpHigh}{
  {m_t \over \Gamma} {d\Gamma \over dE_u} \propto
   {E_u^2 \over m_t^2} 
   \left[ 1 - {2E_u \over m_t} - (h_{\rm UV}-1) {E_u^2 \over m_t^2}
     \right]^{\Delta_\phi^{\rm UV}-2} 
     \theta\left( {m_t \over 1 + \sqrt{h_{UV}}} - E_u \right) \,.
 }
The main qualitative feature of \eno{UpHigh} is that the up quark energy spectrum stops at an energy $m_t / (1 + \sqrt{h_{\rm UV}})$, lower than the usual $m_t / 2$.  This is a direct consequence of the narrower momentum-space light-cone in which the dominant ultraviolet contribution to the unparticle Green's function lies.  To appreciate this point without going through amplitudes explicitly, note that $\omega_{\cal U} = m_t - E_u$ and $|\vec{k}_{\cal U}| = |\vec{k}_u| = E_u$, so the condition that $p_\mu^{\cal U} = (-\omega_{\cal U},\vec{k}_{\cal U})$ is UV-timelike becomes, in two equivalent forms,
 \eqn{WhenTimelike}{
  \omega_{\cal U} &\geq \sqrt{h_{\rm UV}} |\vec{k}_{\cal U}|  \cr
  m_t - E_u &\geq \sqrt{h_{\rm UV}} E_u \,.
 }
The second of these is clearly equivalent to $E_u \leq m_t / (1 + \sqrt{h_{\rm UV}})$.  In figure~\ref{PHASECUTOFF} I show how the energy distribution \eno{UpEnergies} interpolates between the infrared limit, where $W(k_{\cal U}) = 1$, and the ultraviolet limit \eno{UpHigh}.  Evidently, there are at least two difficulties in using a ``signal'' of the type I have described to detect the presence of time warp effects: first, the curves with large $m_t L$ could easily be confused with standard kinematics with a lower value of $m_t$; and second, the curves with moderate $m_t L$ may be difficult to distinguish from unparticle effects with $\Delta$ slightly larger than $3$.  An optimal circumstance would be to have large $m_t L$ and an independent determination of $m_t$.  (By $m_t$, I mean now the mass of some heavy particle with a decay like $t \to u + {\cal U}$---it doesn't have to be the top, nor does $u$ have to be an up quark, just some light visible sector particle.)

\section{The Planck brane}
\label{BRANE}

Adding a Planck brane means adding a term to the action:
 \eqn{FullAction}{
  S = S_{\rm bulk} + S_{\rm brane} \,,
 }
where
 \eqn{Sbrane}{
  S_{\rm brane} = {1 \over 16\pi G_5} \int d^4 x \, \sqrt{h}
      {\cal L}_{\rm brane} \,,
 }
and $h_{mn}$ is the induced metric on the brane.  This extra term results in additional terms in the equations of motion which are distributions supported at the position of the brane, say at $r=0$.  For purposes of calculation, it is convenient to think of two mirror-image copies of the bulk geometry separated by the Planck brane.  This can be thought of as the ``upstairs'' picture of a ${\bf Z}_2$ orbifold acting as $r \to -r$.  Such a picture is well-motivated from string theory \cite{Polchinski:1995df,Horava:1995qa,Horava:1996ma,Lukas:1998yy,Lukas:1998tt} and has been widely considered following \cite{Randall:1999vf}.  Indeed, most of the earlier works on asymmetrically warped geometries summarized in section~\ref{INTRODUCTION} employ a Planck brane.  Some of the results of this section are quite standard: for example the junction conditions on the metric components are special cases of more general relations derived in \cite{Binetruy:1999ut}.  However, in contrast to \cite{Csaki:2000dm}, I assign positive parity to the ${\bf R}^{3,1}$ components of $A_\mu$ under the ${\bf Z}_2$ orbifold symmetry and negative parity to $A_r$.

For simplicity, I will work in the following axial gauge throughout this section:
 \eqn{AxialGauge}{
  ds_5^2 = g_{mn} dx^m dx^n + dr^2  \qquad
   A = A_m dx^m  \,,
 }
where $m$ and $n$ run from $0$ to $3$.  The metric coefficients $g_{mn}$, the gauge field components $A_m$, and the scalar $\psi$ may depend on $r$ as well as $x^m$.  It will be convenient to define a unit normal $n_\mu dx^\mu = dr$ to the brane, where, as usual, $\mu$ runs over all five dimensions of the bulk.  Then one may express the induced metric as a $5 \times 5$ tensor as $h_{\mu\nu} = g_{\mu\nu} - n_\mu n_\nu$.  In the following, I will pass freely between four- and five-dimensional forms of tensors on the brane whose components in the $\mu=5$ direction vanish.

The equations of motion resulting from \eno{FullAction} are
 \eqn{BraneEOMs}{
  G_{\mu\nu} &= {1 \over 2} T_{\mu\nu}^{\rm bulk} + 
    {1 \over 2} T_{\mu\nu}^{\rm brane} \delta(r)  \cr
  \nabla_\mu F^{\mu\nu} &= J^\nu_{\rm bulk} +
    J^\nu_{\rm brane} \delta(r)  \cr
  D_\mu D^\mu \psi &= {\partial V \over \partial \psi^*} + 
    j^{\rm brane} \delta(r) \,,
 }
where
 \eqn{BulkSources}{
  T_{\mu\nu}^{\rm bulk} &= 2 D_\mu \psi^* D_\nu \psi -
    |D\psi|^2 g_{\mu\nu} - V g_{\mu\nu} + 
    F_{\mu\alpha} F_\nu{}^\alpha - 
    {1 \over 4} g_{\mu\nu} F_{\alpha\beta}^2  \cr
  J_\mu^{\rm bulk} &= 
    iq (\psi^* \partial_\mu \psi - \psi \partial_\mu \psi^*) + 
    2q^2 A_\mu |\psi|^2  \cr
  D_\mu \psi &\equiv (\partial_\mu - i q A_\mu) \psi
 }
and the brane sources are defined by
 \eqn{BraneSources}{
  \delta S_{\rm brane} = 
    {1 \over 16\pi G_5} 
     \int d^4 x \, \sqrt{h} \left[ {1 \over 2} \delta h^{mn}
      T_{mn}^{\rm brane} - \delta A^m J_m^{\rm brane} - 
      \delta\psi^* j^{\rm brane} - \delta\psi j^{\rm brane,*}
    \right] \,.
 }
Assuming $g_{mn}$, $A_m$ and $\psi$ are smooth functions of $r$ except for jumps in the first derivatives at $r=0$, one can extract the brane source terms by integrating \eno{BraneEOMs} over a small interval around $r=0$: for example,
 \eqn{TbraneInt}{
  T_{\mu\nu}^{\rm brane} = 2 \lim_{\epsilon \to 0+} 
    \int_{-\epsilon}^\epsilon dr \, G_{\mu\nu} \,,
 }
where I've omitted $T_{\mu\nu}^{\rm bulk}$ because it only involves first derivatives, which have no singular part.  One easily finds
 \eqn{CovariantPlanck}{
  T_{\mu\nu}^{\rm brane} &= 
    4 \left[ K_{\mu\nu} - K h_{\mu\nu} \right]_{0^-}  \cr
  J_\mu^{\rm brane} &= -2 \left[ F_{r\mu} \right]_{0^-} \cr
  j^{\rm brane} &= -2 \left[ \partial_r \psi \right]_{0^-} \,,
 }
where in general,
 \eqn{EvalZeroMinus}{
  \left[ f(r) \right]_{0^-} \equiv \lim_{r \to 0^-} f(r) \,,
 }
and the extrinsic curvature is
 \eqn{KmnDef}{
  K_{\mu\nu} = -h_\mu{}^\rho \nabla_\rho n_\nu = -{1 \over 2} 
    {\partial h_{\mu\nu} \over \partial r} \,.
 }
The middle expression in \eno{KmnDef} is the defining equation for $K_{\mu\nu}$, and the last expression is a result of the gauge choice \eno{AxialGauge}.  $K$ denotes the trace $g^{\mu\nu} K_{\mu\nu}$.  $K_{\mu r} = 0$ for all $\mu$, and as a consequence, $T^{\rm brane}_{\mu r} = 0$.

For brevity, let's define
 \eqn{BraneDefs}{
  \epsilon = -T_0^{0,{\rm brane}} \qquad
  p = {1 \over 3} T_i^{i,{\rm brane}} \qquad
  \rho = J_0^{\rm brane} \qquad
  j = j^{\rm brane} \,,
 }
where $i$ runs from $1$ to $3$.  Plugging the ansatz \eno{AdSansatz} into \eno{CovariantPlanck} leads to
 \eqn{PlanckProperties}{
  \epsilon = \left[ 12 A' \right]_{0^-} \qquad
  p = \left[ -12 A' - {2h' \over h} \right]_{0^-} \qquad
  \rho = \left[ -2\Phi' \right]_{0^-} \qquad
  j = \left[ -2\psi' \right]_{0^-} \,,
 }
Thus we see that the Planck brane must carry electric charge under the gauge field $A_\mu$, and its stress tensor must also break Lorentz invariance.\footnote{If I had assigned negative parity to $\Phi$, as in \cite{Csaki:2000dm}, then the boundary condition at the Planck brane would be $\Phi=0$.  This cannot be reconciled with the requirement that $\Phi \to 0$ in the infrared part of the geometry.  Perhaps some variant of the bulk solution could accommodate non-zero $\Phi$ in the infrared.  But it would be impossible to recover Lorentz symmetry in the infrared with both $\Phi$ and $\psi$ non-zero there, because the conserved current $J_\mu = i \psi^* \overleftrightarrow\partial_\mu \psi + 2 q^2 A_\mu \psi^* \psi$ coupled to $A_\mu$ has $J_0 \neq 0$.  If $\psi$ is zero in the infrared, then non-zero $\Phi$ is possible, but there is still some potential difficulty: the gauge field $A = \Phi dt$ is ill-defined as a 1-form at a horizon where $g^{tt} \to \infty$.}

Provided $A'>0$ and $h'>0$, \eno{PlanckProperties} implies that the Planck brane has an equation of state with $w \equiv {p \over \epsilon} < -1$.  This is a violation of the null energy condition---essentially the same one as found in \cite{Csaki:2000dm} in the absence of a charged scalar.  Some such violation was inevitable because of the argument due to \cite{Cline:2001yt} and outlined in section~\ref{BULK}: $h$ cannot have a maximum as long as the null energy condition is obeyed and $A$ is well defined.  To examine this in more detail, consider cutting off the geometry shown in figure~\ref{EXAMPLE} with a Planck brane, not at $r=0$, but at some radius $r=r_*$ which is positive enough that we can use the ultraviolet asymptotics \eno{UVasymp}.  Then I find that
 \eqn{FoundW}{
  w = -1 + {p_1 p_2 \over 3 h_{\rm UV}} e^{-4r_*/L} + \ldots \,,
 }
where $\ldots$ indicates terms that are even more exponentially suppressed at large $r_*$.  We do have $w<-1$ since $p_1 p_2 < 0$; but it is notable that as $r_*$ increases, $w$ gets exponentially close to $-1$.  So, by this measure, we don't have to violate the null energy condition by much.

For the sake of exhibiting a definite construction, let's consider how one might accommodate \eno{PlanckProperties} using gauged phantoms on the brane.  The first step, largely following \cite{ArmendarizPicon:1999rj}, is to assume a brane lagrangian of the form
 \eqn{BraneL}{
  {\cal L}_{\rm brane} = f(X) - V_{\rm brane}(|\psi|) \,,
 }
where
 \eqn{Xdef}{
  X \equiv -|D_m \psi|^2
 }
and $f(X)$ is some smooth function.  The gauge-covariant derivative $D_m$ is the same as the one in \eno{BulkSources}.  Assuming that $D_i\psi = 0$ for $i=1,2,3$ and that $\partial_0 \psi = 0$, one readily obtains
 \eqn[c]{FoundSources}{
  X = -h^{00} q^2 \Phi^2 |\psi|^2 \qquad
  \epsilon = 2X f'(X) - f(X) + V_{\rm brane}(|\psi|) \qquad
   p = f(X) - V_{\rm brane}(|\psi|)  \cr
  \rho = 2 f'(X) q^2 \Phi |\psi|^2 \qquad
  j = h^{00} q^2 \Phi^2 \psi f'(X) + 
    {\partial V_{\rm brane} \over \partial\psi^*} \,.
 }
The expressions \eno{FoundSources} satisfy a first law constraint:
 \eqn{FirstLawBrane}{
  \epsilon + p + h^{00} \rho \Phi = 0 \,.
 }
The same first law constraint holds for the bulk relations \eno{PlanckProperties} once one imposes
 \eqn{ExtremalID}{
  h' = e^{-2A} \Phi \Phi' \,,
 }
which is what one gets by demanding that the Noether charge \eno{NoetherChargeGeneral} vanishes.\footnote{In section~\ref{INTRODUCTION} I remarked that a simple way to construct a time warp geometry---that is, a geometry where time has a different warp factor from space, but there is no temperature or entropy associated with the extra-dimensional geometry---is to start with a black brane geometry and cut it off above the horizon, as done for example in \cite{Csaki:2000dm,Cline:2003xy}.  With the exception of extremal Reissner-Nordstr\"om, the bulk geometries considered in these works do not obey the extremality condition \eno{ExtremalID}.  So the bulk geometry will demand an equation of state on the brane that is different from \eno{FirstLawBrane}.  The difference is essentially a $Ts$ term, where $T$ is the temperature and $s$ is the entropy density that the horizon would have had if it were present.  It seems to me a non-trivial difficulty to construct a theory on the brane that will accommodate the equation of state that the bulk demands without involving non-zero entropy and temperature.  For this reason, it is not clear that the infrared cutoff is a satisfactory construction.

The Planck brane in \cite{Csaki:2000dm} is required to have an equation of state $\epsilon + p < 0$, just as I found in \eno{FoundW}.  It was not demonstrated, however, that such an equation of state actually arises from the dynamics on the brane, decoupled as it is from the $U(1)$ gauge field when $A_m$ has odd parity.  This is in contrast to \eno{FoundSources}-\eno{FirstLawBrane}, where the equation of state is seen to arise explicitly from the gauged phantom construction, and to be a consequence more generally of the first law of thermodynamics.}

Suppose we start with a bulk solution with $Q=0$ and want to tailor the functions $f(X)$ and $V_{\rm brane}(|\psi|)$ so that the Planck brane ``fits'' onto the bulk solution at a specified radius, $r=r_*$.  ``Fitting'' means that \eno{PlanckProperties} and \eno{FoundSources} are consistent at the specified radius.  In light of \eno{FirstLawBrane}-\eno{ExtremalID}, we need only demand that the last three equations of  \eno{PlanckProperties} are consistent with the last three of \eno{FoundSources}.  From these requirements, we can extract the following conditions:
 \eqn{fConditions}{
  X &= \left[ {e^{-2A} q^2 \Phi^2 |\psi|^2 \over h} \right]_{r_*^-}
     = {q^2 e^{-2a_1} p_1^2 s_2^2 \over h_{\rm UV}}
       e^{-2 (1+\Delta_\psi^{\rm UV}) r_*/L} + \ldots \cr
  f(X) - V_{\rm brane}(|\psi|) &= 
    \left[ -12 A' - {2h' \over h} \right]_{r_*^-} 
   = -{12 \over L} + {p_1 p_2 \over h_{\rm UV} L} e^{-4r_*/L}
     + \ldots  \cr
  X f'(X) &= \left[ -{h' \over h} \right]_{r_*^-} 
   = {2 p_1 p_2 \over h_{\rm UV} L} e^{-4r_*/L} + \ldots  \cr
  \psi^* {\partial V_{\rm brane} \over \partial\psi^*} &= 
    \left[ -{h' \over h} - 2 \psi^* \psi' 
     \right]_{r_*^-} = {2 p_1 p_2 \over h_{\rm UV} L}
       e^{-4r_*/L} + 
     {2 s_2^2 \Delta_\psi^{\rm UV} \over L} 
      e^{-2\Delta_\psi^{\rm UV} r_*/L} + \ldots \,,
 }
where $\ldots$ indicates terms which are subleading in the UV expansions \eno{UVasymp} compared to the ones shown.  What the equations \eno{fConditions} mean is that $f(X)-V_{\rm brane}(|\psi|)$, $X f'(X)$, and $\psi^* \partial V_{\rm brane} / \partial\psi^*$ are required to take on the values indicated for the particular value of $X$ indicated, and for $\psi$ evaluated on the brane.  These conditions say nothing about the global shape of $f$ as a function of $X$ or of $V_{\rm brane}$ as a function of $|\psi|$.  But the third equation in \eno{fConditions} indicates that $f'(X) < 0$.  This is the characteristic feature of phantoms, and it could also have been anticipated by noting that $w = -1 + 2X f'(X) / \epsilon < 0$.

Having $f'(X) < 0$ raises questions about stability and/or unitarity: see for example \cite{Cline:2003gs}.  However, it seems likely that a combination of positive $f''(X)$, positive curvature for $V_{\rm brane}(|\psi|)$, and augmentation of ${\cal L}_{\rm brane}$ by appropriate higher derivative terms, as in \cite{Creminelli:2006xe}, would lead to a stable construction.  There is no guarantee, of course, that the requisite properties of ${\cal L}_{\rm brane}$ are reasonable, in the sense of being in the ballpark of what one might obtain from studying an actual string theory compactification.  Gauged phantoms are probably not the only option for an appropriate Planck brane construction.  Certainly, other ways of violating the null energy condition have been discussed, as has the link between null energy violations and superluminal motion.  For a few points of entry into the large and diverse literature on these topics, see \cite{Alcubierre:1994tu,ArmendarizPicon:2000ah,Sahni:2002dx,Dubovsky:2005xd,Buchbinder:2007ad,Marvel:2008uh}.

\section{Gravitons}
\label{GRAVITON}

Having passed from an asymptotically $AdS_5$ geometry to a geometry with a UV cutoff---through the admittedly {\it ad hoc} introduction of gauged phantoms on the Planck brane---the next question to ask is how gravity works from a four-dimensional perspective.  An all-but-necessary condition for physically reasonable gravity is the existence of a spin two particle which propagates at the infrared speed of light, at least when it is not too energetic.  The purpose of this section is to ask what we have to do to get such a four-dimensional graviton.  The answer is simple to state: we must add a wrong-sign Einstein-Hilbert term to the lagrangian on the brane, so that it reads
 \eqn{CompleteLbrane}{
  {\cal L}_{\rm brane} = \eta {}^{(4)}R + f(X) - 
    V_{\rm brane}(|\psi|) \,,
 }
where ${}^{(4)}R$ is the four-dimensional Ricci scalar.  By wrong-sign I mean that $\eta$ has to be negative, so this somewhat different from the proposal of \cite{Dvali:2000hr}.  Moreover, $\eta$ needs to be tuned to a certain value, close to $-L$, in order to make the four-dimensional graviton appear with the desired infrared dispersion relation, $\omega = |\vec{k}|$.

A plane wave of gravitons moving in the $x^1$ direction and polarized in the $x^2$-$x^3$ direction can be described by the following perturbation of the metric \eno{AdSansatz}:
 \eqn{PerturbedMetric}{
  ds_5^2 = e^{2A(r)} \left[ -h(r) dt^2 + d\vec{x}^2 + 
    2 e_{23}(r) \cos(\omega t - kx) dx^2 dx^3 \right] + 
      dr^2 \,,
 }
where as in \eno{AxialGauge} I use an axial gauge.  Treating the perturbation to linear order, and accounting for the Planck brane action \eno{CompleteLbrane}, one finds that the perturbation obeys the equation
 \eqn{ETwoThree}{
  e_{23}'' + \left[ 4A'+{h' \over 2h} \right] e_{23}' + 
    e^{-2A} \left[ 1 + \eta \delta(r) \right] 
    \left( {\omega^2 \over h} - k^2 \right) e_{23} = 0 \,,
 }
where primes denote $d/dr$ and I have assumed, for now, that the Planck brane is at $r=0$.  If $h=1$ and $\eta=0$, as in \cite{Randall:1999vf}, then the solution describing a graviton is $e_{23}=1$, with the on-shell requirement $\omega^2 = k^2$.

To treat the general case with $h \neq 1$ and $\eta \neq 0$, one must first extract the correct boundary condition at the Planck brane.  As with the boundary conditions \eno{CovariantPlanck}, this is done by integrating the equation of motion \eno{ETwoThree} over a small region including $r=0$.  The result is
 \eqn{BDgrav}{
  2 \left[ e_{23}' \right]_{0^-} = 
    \eta \left[ e^{-2A} \left( {\omega^2 \over h} - 
      k^2 \right) e_{23} \right]_{0^-} \,.
 }
My aim is to find out what $\eta$ we should choose in order to have a graviton that travels at the infrared speed of light.  That is, I simply require that the on-shell condition is $\omega^2 = k^2$.  Imposing this restriction, one can solve \eno{BDgrav} for $\eta$ to find
 \eqn{FoundEta}{
  \eta = -{2 \over \omega^2} \left[ e_{23}' / e_{23} \over 
    e^{-2A} \left( 1 - {1 \over h} \right) \right]_{0^-} \,.
 }
The correct boundary condition in the infrared is that $e_{23}$ approaches $1$: its infrared behavior is then the same as the graviton found in \cite{Randall:1999vf}.  It is very likely that the value of $\eta$ one extracts from \eno{FoundEta} depends on $\omega$, which means that we can't make gravitons with arbitrary energy all travel with the infrared speed of light.  We must be satisfied instead with the choice of $\eta$ that makes gravitons with $\omega L \ll 1$ IR-lightlike.\footnote{Perhaps by introducing on the Planck brane a series of higher derivative terms in the curvature one could obtain a dispersion relation for four-dimensional gravitons of the form $\omega^2 = k^2$ with corrections suppressed by any desired positive even integer power of $\omega L$.  Even if this is possible, each successive term would presumably have a coefficient that needs to be tuned to a particular value.}  To calculate this value, let's solve \eno{ETwoThree} with $\omega^2 = k^2$ perturbatively in small $\omega$ by expanding
 \eqn{ExpandE}{
  e_{23} = \phi_0 + \omega^2 \phi_2 + \omega^4 \phi_4 + \ldots \,.
 }
At zeroth order in $\omega$, the equation of motion \eno{ETwoThree} becomes
 \eqn{ZeroethOrderPhi}{
  \phi_0'' + \left[ 4A' + {h' \over 2h} \right] \phi_0' = 0 \,,
 }
and the Planck brane boundary condition \eno{BDgrav} becomes $\big[\phi_0'\big]_{0^-} = 0$.  The solution of \eno{ZeroethOrderPhi} satisfying this boundary condition is $\phi_0 = 1$.  At the next order in $\omega$, the equation \eno{ETwoThree} becomes
 \eqn{LeadingCorrection}{
  \phi_2'' + \left[ 4A' + {h' \over 2h} \right] \phi_2' = 
    e^{-2A} \left( 1 - {1 \over h} \right) \,.
 }
away from $r=0$.  This equation is readily integrated.  Plugging the result into \eno{FoundEta}, one sees that
 \eqn{EtaResult}{
  \eta = - \left[ {2 \over e^{2A} \sqrt{h} \left( 1 - {1 \over h}
    \right)} \right]_{0^-} 
    \int_{-\infty}^0 dr \, e^{2A} \sqrt{h} \left( 1 - {1 \over h}
      \right) \,.
 }
If we wish the Planck brane to be at some radius other than $0$, say $r=r_*$, then we should evaluate the prefactor in \eno{EtaResult} at $r_*^-$ and change the upper limit of integration to $r_*$.  Assuming that $r_*$ is well into the ultraviolet region, where $A \approx r/L + a_1 + \ldots$ and $h \approx h_{\rm UV}$, one sees that the dominant contribution to the integral in \eno{EtaResult} comes from the ultraviolet region.  Plugging these approximate forms into \eno{EtaResult}, one finds
 \eqn{ApproxEta}{
  \eta \approx -L \,,
 }
as claimed at the beginning of this section.  Using the numerical solution exhibited in figure~\ref{EXAMPLE}, and setting $r_* = 3L$, one finds $-\eta/L \approx 0.998$.

\section{Discussion}
\label{DISCUSSION}

The main idea of a time warp compactification is that particles could travel faster than the observed speed of light if they can propagate through some region of an extra-dimensional spacetime where time is gravitationally blue-shifted.  There are some serious obstacles to realizing this idea.  First, there are the constraints from the null energy condition, as discussed in sections~\ref{INTRODUCTION} and~\ref{BULK}, extending the arguments of \cite{Cline:2001yt}.  Second, we have seen in section~\ref{GRAVITON} that it is non-trivial to obtain a four-dimensional, spin-two, infrared-massless graviton.  And third, we must remember that there are stringent experimental limits on violations of Lorentz invariance.

As an example of the experimental limits, consider for example vacuum Cerenkov effects, as discussed in \cite{Coleman:1998ti,Jacobson:2001tu} for the case of photon emission, and in \cite{Moore:2001bv,Cline:2003xy} for the case of graviton emission.  For electrons, the bound quoted in \cite{Coleman:1998ti} is $(c_e-c_\gamma)/c_\gamma \lsim 5 \times 10^{-13}$, where $c_\gamma$ is the speed of light and $c_e$ is the limit on the speed of electrons.  Observation of primary protons in cosmic rays with energies of up to $10^{20}\,{\rm eV}$ indicates an even tighter bound for protons, $(c_p-c_\gamma)/c_\gamma \lsim 10^{-23}$.  Can we possibly get away with a construction like the one exhibited in figure~\ref{EXAMPLE}, in which which maximum speed varies across the extra dimension by a factor of $1.7$?

The approach considered in section~\ref{CORRELATORS}, where Lorentz violation originates entirely from a strongly coupled unparticle sector, may provide a way to evade the existing constraints on Lorentz violations with unparticle Green's functions modified on a scale $1/L$ comparable to the $\rm TeV$ scale.  Such modifications are in principle discoverable at the LHC, for example through the unusual kinematic constraints illustrated in figure~\ref{PHASECUTOFF}.  It should be noted, however, that I did not try to quantify the extent to which virtual unparticles might communicate Lorentz violations into visible sector propagators and couplings.

One could also entertain the possibility that the visible sector itself is dual to a time warp geometry.  It's hard to see how to accommodate this idea without setting the scale $1/L$ where Green's functions are modified quite high.  The idea that boost invariance is an emergent infrared symmetry is an old one, dating back at least to \cite{Chadha:1982qq}.  But the experimental constraints on high-energy modifications of dispersion relations are pretty tight: see for example \cite{Jacobson:2001tu}.  Nevertheless, it seems to me significant that one can start with a generally covariant theory, spontaneously break Lorentz invariance at a high scale and, through quite an explicit extra-dimensional construction, recover it in the infrared.

A notable feature of time warp geometries is that the speed of light, $h(r)$, varies exponentially slowly both in the infrared and the ultraviolet, as a function of proper distance $r$ in the fifth dimension: see the asymptotic expressions~\eno{IRasymph} and~\eno{UVasymph}.  This suggests the possibility of generating extremely small but non-zero differences between the maximum attainable velocities of different Standard Model particles by having them propagate on branes at significantly different locations deep in the infrared part of the geometry.  If a particle comes from a string stretched between two branes, then its maximum attainable velocity is dictated by the brane that is deeper in the infrared, as explained in \cite{Peeters:2006iu,Herzog:2006gh,Liu:2006nn,Chernicoff:2006hi,Mateos:2007vn,Ejaz:2007hg,Argyres:2008eg,CasalderreySolana:2008ne} in a finite-temperature setting.  But can one arrange for appropriate couplings among particles on branes separated in such a way?

The time warp geometry I constructed is just one example based on the simplest possible lagrangian.  A diverse collection of other solutions, with remarkably variable $h_{\rm UV}$, can be found just by varying the parameters in this lagrangian.  See for example \cite{Gubser:2008pf}, where, apparently, an exponentially large $h_{\rm UV}$ was achieved by varying $q$ over a modest range.  Although the lagrangian I use is not taken directly from a string theory construction, the ingredients are generic enough that I certainly expect it, or something with qualitatively similar solutions, can be embedded into string theory constructions.  More generally, one could try to support a time warp geometry with different combinations of fields.  Interesting field combinations include $B_p = dt \wedge \omega_{p-1}$, where $B_p$ is a $p$-form gauge potential and $\omega_{p-1}$ is a $(p-1)$-form on the extra dimensions; or perhaps some fermion bilinear like $\bar\psi \gamma^1 \gamma^2 \gamma^3 \psi$.  In most circumstances, I expect that a violation of the null energy condition would be necessary in order to achieve a static geometry.  Orientifold planes violate the null energy condition, but in a way that allows overall, boost-invariant warping of the geometry, not time warping.  Recall that the constraints on $h$ came from considering the combination $R^0_0 - R^1_1$, where the $1$ direction is one of the usual dimensions of space; but orientifold planes extended over ${\bf R}^{3,1}$ are, by themselves, Lorentz-invariant, so they do not contribute to $R^0_0 - R^1_1$.  Perhaps one could arrange for some Casimir effect to generate $w < -1$ on the Planck brane; or perhaps some higher derivative terms in the bulk would loosen the constraints.

I have left a number of issues unexplored.  Here is a partial list:
 \begin{itemize}
  \item Although I have speculated that a stable configuration, including a violation of the null energy condition on the Planck brane, could be achieved, I have by no means demonstrated this explicitly.  To demonstrate stability, one would presumably have to start by studying the coupled perturbations of all the bulk and brane fields---already a non-trivial problem.
  \item The key feature of the bulk geometry is the $SO(3,1)$ symmetry that emerges in the infrared.  One could plausibly arrange other emergent symmetries.  For example, if a scalar runs in the infrared to an extremum of a potential where some gauge symmetry is restored, then this symmetry could be fairly described as emergent.  Without some cutoff like the Planck brane, a gauge symmetry in the bulk corresponds to a global symmetry on the boundary.  It might be instructive to see how (approximate) gauge invariance results from an appropriate cutoff.
  \item Although I have shown that one can contrive to have a spin-two graviton which propagates at the infrared speed of light, I did not show that the low-energy physics includes standard Einstein gravity.  I expect that perturbations of non-transverse-traceless components of the metric mix with other fields, and it is a matter of calculation to find out how they affect low-energy four-dimensional physics.
  \item If there really is more-or-less standard gravity in the four-dimensional effective lagrangian, then black hole physics provides another interesting set of questions.  Can one see part way inside a black hole with particles that propagate faster than the infrared speed of light?
  \item Time warp geometries with more than one extra dimension might offer some novel possibilities.  For example, one might try to evade the null energy constraints by considering an extra-dimensional which is non-compact due to a finite-volume ``spike'' along which $h$ increases without ever reaching a maximum.
  \item I considered only the simplest interface between time warps and unparticle physics.  A host of related calculations could be revisited, either with some rough constraints in mind (like the ultraviolet ``un-shell'' condition $\omega_{\cal U}^2/h_{\rm UV} - \vec{k}_{\cal U}^2 \geq 0$), or with some more precise evaluations of Green's functions in hand.
  \item The conditions \eno{fConditions} on $X$, $f(X)$, and $V_{\rm brane}(|\psi|)$ at the Planck brane involve exponentially small quantities and, most likely, some fine-tuning; also the choice of $\eta$ in section~\ref{GRAVITON} seems to be a fine-tuning.  Fine-tuning might be hard to avoid altogether, but I would not be surprised if the example I studied explicitly is far from the most natural time warp construction.
 \end{itemize}
I hope to report on some of these issues in the future.

\section*{Acknowledgments}

I thank J.~Khoury, A.~Nellore, S.~Pufu, F.~Rocha, L.~Senatore, P.~Steinhardt, and A.~Yarom for useful discussions.  This work was supported in part by the Department of Energy under Grant No.\ DE-FG02-91ER40671 and by the NSF under award number PHY-0652782.

\clearpage
\appendix

\section{More on two-point correlators}
\label{GREENS}

The asymptotic forms of $f_k(r)$ are straightforward generalizations of \eno{AdSGF}:
 \eqn{fpIR}{
  f^{\rm IR}_k(r) = \left\{ \seqalign{\span\TR & \qquad\span\TT}{
    e^{-2A} H_{\Delta_\phi^{\rm IR}-2}(L_{\rm IR}
      \sqrt{\omega^2-\vec{k}^2} e^{-A}) &
     for $\omega^2 > \vec{k}^2$  \cr
    e^{-2A} K_{\Delta_\phi^{\rm IR}-2}(L_{\rm IR}
      \sqrt{\vec{k}^2-\omega^2} e^{-A}) &
     for $\omega^2 < \vec{k}^2$
  } \right.
 }
in the infrared, and 
 \eqn{fpUV}{
  f_k^{\rm UV}(r) = \left\{ \seqalign{\span\TR & \qquad\span\TT}{
   \eqalign{ 
    a_k e^{-2A} & J_{-\Delta_\phi^{\rm UV}+2}(L
      \sqrt{\omega^2/h_{\rm UV}-\vec{k}^2} e^{-A})  \cr
     &\qquad{} + 
    b_k e^{-2A} J_{\Delta_\phi^{\rm UV}-2}(L
      \sqrt{\omega^2/h_{\rm UV}-\vec{k}^2} e^{-A})
   } & for $\displaystyle{\omega^2 \over h_{\rm UV}} > \vec{k}^2$  
     \cr\noalign{\vskip2\jot}
   \eqalign{ 
    a_k e^{-2A} & I_{-\Delta_\phi^{\rm UV}+2}(L
      \sqrt{\vec{k}^2-\omega^2/h_{\rm UV}} e^{-A})  \cr
     &\qquad{} + 
    b_k e^{-2A} I_{\Delta_\phi^{\rm UV}-2}(L
      \sqrt{\vec{k}^2-\omega^2/h_{\rm UV}} e^{-A})
   } & for $\displaystyle{\omega^2 \over h_{\rm UV}} < \vec{k}^2$
  } \right.
 }
in the ultraviolet, where $a_k$ and $b_k$ are coefficients that have to be determined by solving the full equations of motion.  In terms of these coefficients, the full Green's function is
 \eqn{FullGF}{
  G_F(k) = -2K_{\cal O}^{\rm UV} \sqrt{h_{\rm UV}}
    {\Gamma(3 - \Delta_\phi^{\rm UV}) \over
      \Gamma(\Delta_\phi^{\rm UV} - 2)}
    \left| {\omega^2 \over h_{\rm UV}} - \vec{k}^2 
     \right|^{\Delta_\phi^{\rm UV}-2} {b_k \over a_k} \,.
 }
In evaluating $a_k$ and $b_k$, one must impose the same infrared boundary conditions as discussed following \eno{AdSGF}.  To see that this agrees with the prescription \eno{GpEval} for infrared-timelike momenta (with $\Delta \to \Delta_\phi^{\rm UV}$ and $K_{\cal O} \to K_{\cal O}^{\rm UV}$), first recall that
 \eqn{BesselAsymp}{
  J_\nu(z) = \left( {z \over 2} \right)^\nu {1 \over \Gamma(\nu+1)}
    \left[ 1 + O(z^2) \right] \,,
 }
and that $I_\nu(z)$ has the same leading order expansion for small $z$.  Combining \eno{BesselAsymp} and \eno{fpUV}, one can show that
 \eqn{FUV}{
  {\cal F}_k = -{2 \sqrt{h_{\rm UV}} 
   \sin \pi \Delta_\phi^{\rm UV} \over \pi}
    \Im\{ a_k^* b_k \} \,,
 }
regardless of the sign of $\omega^2 / h_{\rm UV} - \vec{k}^2$.  The result \eno{FUV}, together with the identity
 \eqn{GammaIdent}{
  \Gamma(\nu) \Gamma(1-\nu) = {\pi \over \sin \pi\nu} \,,
 }
is already enough to demonstrate that taking the imaginary part of \eno{FullGF} leads back to \eno{AdSGF}.

The conserved flux can also be evaluated in the infrared:
 \eqn{FIR}{
  {\cal F}_k = -{2 \over \pi} {L \over L_{\rm IR}}
    \theta(\omega^2 - \vec{k}^2) \,,
 }
using the asymptotic form of Hankel functions for large values of the argument.  Combining \eno{GpEval} and \eno{FIR}, one obtains
 \eqn{ImGF}{
  \Im G_F(k) = -{2K_{\cal O}^{\rm UV} \over \pi} {L \over L_{\rm IR}}
    {\Gamma(3-\Delta_\phi^{\rm UV})^2 \over |a_p|^2}
    \left| {\omega^2 \over h_{\rm UV}} - \vec{p}^2
      \right|^{\Delta_\phi^{\rm UV}-2}
   \theta(\omega^2-\vec{p}^2) \,.
 }
The most numerically stable way to evaluate $\Im G_F(k)$ is to use \eno{ImGF}, because only $a_k$ needs to be determined numerically.

\clearpage
\bibliographystyle{ssg}
\bibliography{time}
\end{document}